\documentclass{vow2008}

\usepackage{graphicx}

\title{From 1000\,AU to 1000\,pc:
high proper-motion stars in the solar neighbourhood, radio sources in the
$\sigma$~Orionis cluster, and new X-ray stars surrounding Alnilam}  
\author{Jos\'e A. Caballero}
\affil{Departamento de Astrof\'{\i}sica y Ciencias de la Atm\'osfera,
Facultad de F\'{\i}sica, Universidad Complutense de Madrid, 28040
Madrid, Spain. E-mail: {\tt caballero@astrax.fis.ucm.es}}

\begin{document}

\maketitle


\begin{abstract}
The Virtual Observatory is useful.
I~summarise some of my works where I~extensively use the Aladin sky atlas.
Topics cover from the search and common proper motion confirmation of very
low-mass stars and brown dwarfs in wide ($\rho >$ 1000\,AU) binaries and
multiple systems, to the identification and characterisation of stellar and
substellar populations in young open clusters and OB associations at
heliocentric distances of up to 1000\,pc.
I~present three practical examples of what one can do with Aladin in {\em one
morning}:  
a fruitful proper-motion search of objects with available $ugrizJHK_{\rm s}$
photometry, an analysis of the 21\,cm radio sources towards the young
$\sigma$~Orionis cluster, and a novel study of X-ray young stars surrounding
Alnilam in the Orion Belt.
\end{abstract}


\section{Introduction}

What is the Virtual Observatory (VO)?
One may pompously answer that it is {\em ``an international astronomical
community-based initiative that aims to allow seamless access to distributed
astronomical resources, and to provide the necessary tools to analyse data and
produce scientifically relevant results\footnote{\tt
http://www.ivoa.net, http://www.euro-vo.org}.''}  
Another one may be sharper and more optimistic, and answer that it is {\em ``the
Universe in your computer''}.
The use of available astronomical data archives of space and ground-based
observatories and sky survey databases is widespread in Astronomy. 
As an example, to date (2008 Jan), the SAO/NASA Astrophysics Data System
lists almost 900 refereed papers with the title words ``Sloan Digital Sky
Survey'' (SDSS). 
{\em Hipparcos} parallaxes, digitisations of the Palomar Observatory Sky Survey
(POSS), and near-infrared magnitudes from the Two-Micron All Sky Survey (2MASS)
are also used very often. 
However, surveys, missions, and consortia like them are not the VO, but data
{\em providers} that follow common standards given by the~VO.

What I know is what is {\em not} VO.
It is not applying for telescope time, travelling to a remote observatory,
having sleepless nights, nor reducing and calibrating data when back to office. 
Thus, being a ``virtual observer'' is a cheaper, more ecological, and less
time-wasting way of doing Astronomy (how many telescope nights have you lost due
to bad weather? how many hours have you spent at the airport?).
Obviously, the VO will not kill the classical observing mode.
On the contrary, VO is the perfect complement to telescope facilities of all
sizes.

In this proceeding, I~alternate an ``egocentric'' summary of my VO-related
publications and three basic, practical examples of what one can do with my
favourite VO tool: the Aladin sky atlas (Bonnarel et~al. 2000).
The common denominator of the examples is that they can be fully accomplished in
just {\em a few hours} (less than a working day) and are scientifically
interesting.

\section{Proper-motion surveys}

\begin{table*}
  \begin{center}
    \caption{Some remarkable objects with proper motion $\mu >$ 
    100\,mas\,a$^{-1}$ in the SU2 survey area.}\vspace{1em}
    \renewcommand{\arraystretch}{1.2}
    \begin{tabular}[h]{l cc cc cc l}
      \hline
Name		& $\alpha$      & $\delta$      & $\mu_\alpha \cos{\delta}$& $\mu_\delta$       & $g$		      	& $J$		      	& Reference		\\
		& (J2000)       & (J2000)       & [mas\,a$^{-1}$]       & [mas\,a$^{-1}$]       & [mag]		      	& [mag]		      	& 			\\
      \hline
Albus 2		& 12 00 07.04 	& +30 27 51.5	& --78			& --92			& 16.626$\pm$0.005	& 14.238$\pm$0.030	& {\em this work}	\\ 
Ruber 3		& 12 01 31.39 	& +29 58 49.8	& --62			& --96			& 16.461$\pm$0.004	& 12.147$\pm$0.019	& {\em this work}	\\ 
Albus 3		& 12 02 22.43 	& +30 10 14.1	& +30			& --98			& 16.176$\pm$0.003	& 14.300$\pm$0.028	& {\em this work}	\\ 
LP 320--47	& 12 03 36.27 	& +29 50 00.6	& +38			& --256			& 18.463$\pm$0.007	& 14.935$\pm$0.038	& Luyten 1979		\\ 
GD 147		& 12 04 10.80 	& +30 25 18.4	& --32			& --100			& 16.502$\pm$0.004	& 14.942$\pm$0.047	& Giclas et~al. 1965	\\ 
      \hline
      \end{tabular}
    \label{tab:SU2}
  \end{center}
\end{table*}

When I~serendipitously discovered Koenigstuhl~1~AB (K\"o\,1\,AB; Caballero
2007a), only one very low-mass binary with a projected physical separation $s$
larger than 100\,AU was known, DE0551--44~AB (Bill\`eres et~al. 2005).
With $s \approx$ 220\,AU and a total mass ${\mathcal M}_{\rm A} + {\mathcal
M}_{\rm B} <$ 0.2\,$M_\odot$, DE0551--44~AB represented a challenge to many
low-mass star formation scenarios. 
The projected physical separation between the two components in K\"o\,1\,AB,
although having roughly the same total mass as DE0551--44~AB, was one order of
magnitude larger, $s$ = 1800$\pm$170\,AU, which made it to be by far the widest
low-mass binary.  
Two years later, K\"o\,1\,AB has been surpassed only by 2M0126--50~AB ($s$ =
5100$\pm$400\,AU; Artigau et~al. 2007).
The two components in K\"o\,1\,AB, LEHPM~494 and DE0021--42, were previously
known as late-type, high proper-motion stars, but had not been associated 
because of their faintness and relatively large angular separation of $\rho
\sim$ 1.3\,arcmin.
I~identified K\"o\,1\,AB as a binary candidate during a routine visual
inspection with Aladin and confirmed their common proper-motion with public data
(2MASS, DENIS, SuperCOSMOS digitisations of UK Schmidt plates).
New telescope observations were unnecessary.

In Caballero (2007c), I~extended the search for low-mass stars and brown dwarfs
in wide binaries and multiple systems and measured for the first time the common
proper motion of two new wide systems containing very low-mass components,
Koenigstuhl~2~AB and 3~A-BC.
The two of them are among the widest systems in their respective classes (e.g.
the faintest component in K\"o\,3\,A-BC is the most separated L dwarf to its 
primary, $s$ = 11900$\pm$300\,AU).
In addition, I~determined the frequency of field wide multiples with late-type
components and measured for the first time the proper motions of 62 field stars
and brown dwarfs with spectral types $>$M5.0V.
All the presented results came from on-line catalogues and archival images,
without complementary observations.

New proper-motion surveys aimed at the detection and characterisation of wide
binaries are on-going in collaboration (Caballero et~al. 2008b) or alone
(Caballero, in~prep.).
The basics of these surveys are the extensive use of proper-motion catalogues
(USNO-B1, Tycho-2) with the Aladin sky atlas and will be described in detail
elsewhere.

\subsection{SU2 = SDSS + USNO-B1 + 2MASS}
\label{section:su2}

\begin{figure*}
\centering
\includegraphics[width=0.49\linewidth]{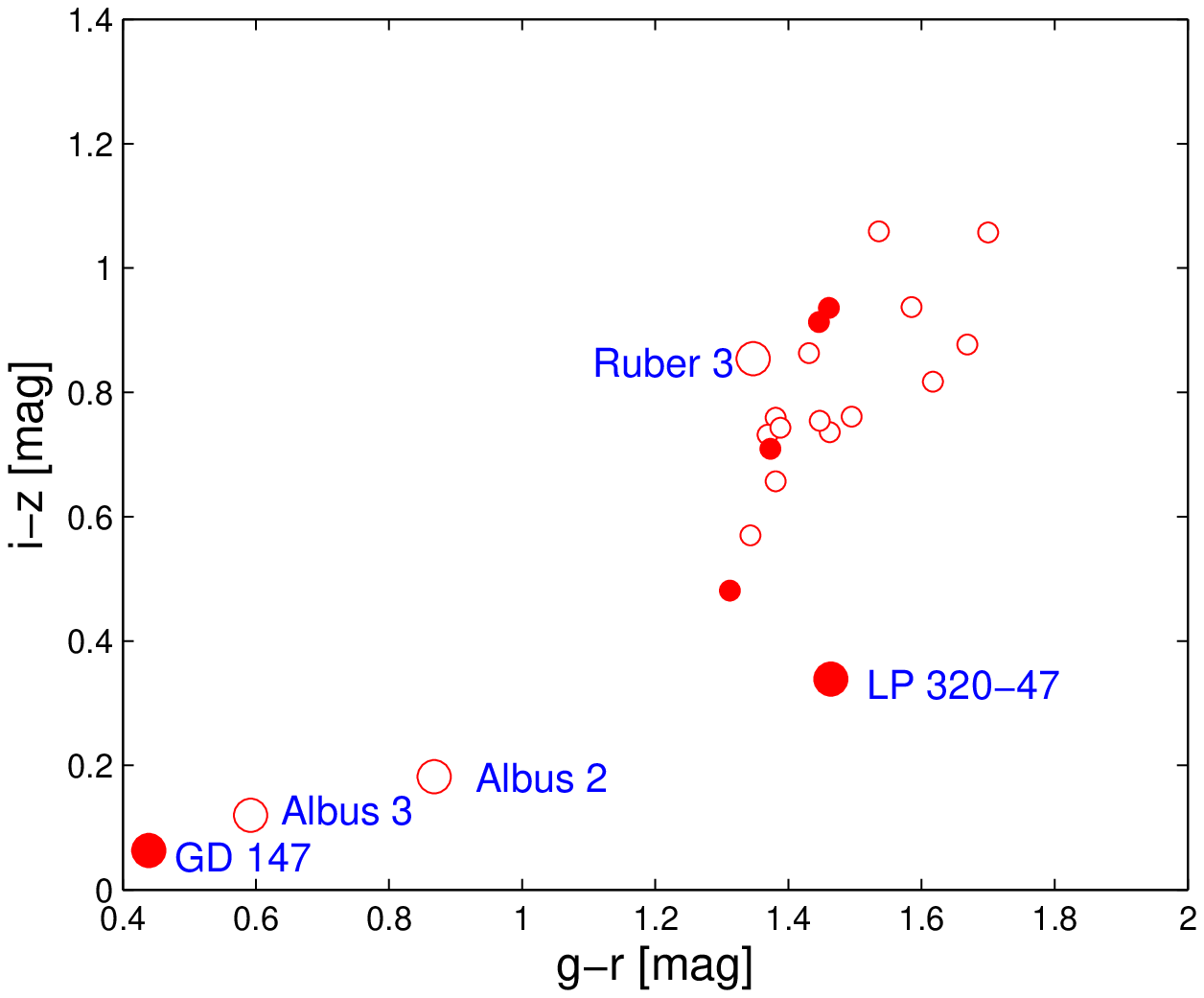}
\includegraphics[width=0.49\linewidth]{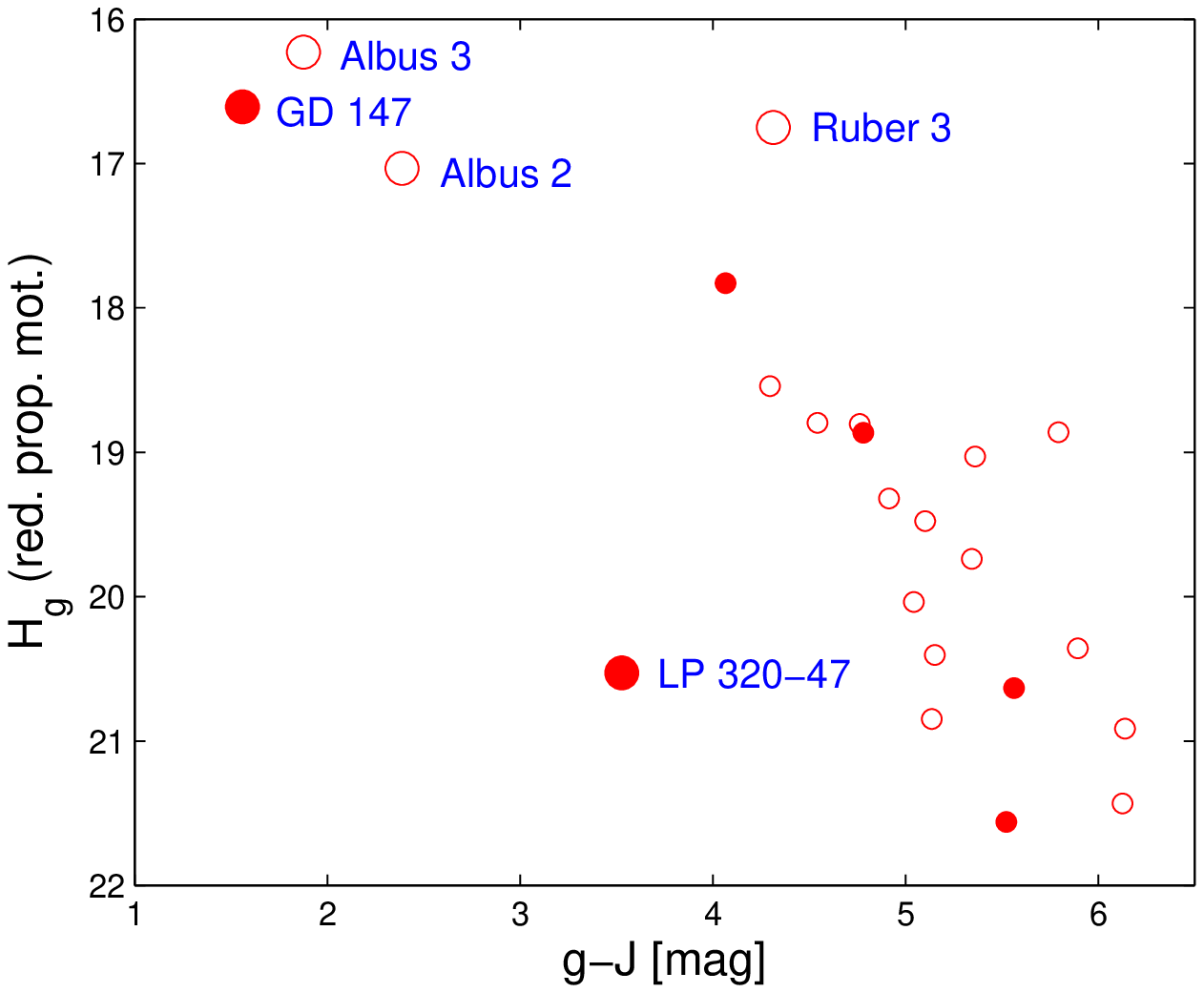}
\caption{Colour-colour ($i-z$ vs. $g-r$, {\em right}) and reduced proper
motion-colour ($H_g$ vs. $g-J$, {\em right}) diagrams of the 23 objects with
proper motion $\mu >$ 100\,mas\,a$^{-1}$ in the SU2 survey area that do not
saturate in the SDSS images.  
Filled circles: objects reported in the literature;
open circles: new objects;
big, labelled circles: remarkable objects in Table~\ref{tab:SU2}.
\label{fig:figure_su2}}
\end{figure*}

The first practical example is a pilot Aladin-based search for high-proper
motion objects. 
One of the disadvantages of most proper motion surveys is the poor follow-up
characterisation of the identified objects.
Classical surveys have provided tentative spectral types based on colours and
magnitudes from photographic plates (e.g. Giclas et~al. 1971; Luyten 1979). 
Low-resolution spectroscopy is available for a tiny fraction of the numerous
high-proper motion dwarf, subdwarf, and white dwarf candidates identified so
far. 
The combination of $V$-band magnitudes estimated from photographic $B_J$ and
$R_F$ magnitudes and near-infrared magnitudes from 2MASS as in Salim \& Gould
(2003) or L\'epine \& Shara (2005) helps a lot in the classification of these
objects, but it is still insuficient.
One way of avoiding this handicap is surveying areas where there exists plenty
of accurate multiband photometry.

I~give the name SU2\footnote{Do not mistake with the special unitary group of
degree 2, SU(2).} to this pilot survey because I~use astro-photometric data from
the SDSS, USNO-B1, and 2MASS catalogues.
Sources in the three catalogues have tabulated coordinates, proper motions, and
magnitudes in the $ugrizJHK_{\rm s}$ bands (apart from the photographic $B_J R_F
I_N$ bands). 
Given the nature of this publication, I~will be ``pedagogical'' in the
description of the SU2 survey.
What follows has done with Aladin~v4, but can be done in the same way with the
most recent release Aladin~v5.015.
The survey has been carried out in a 1\,deg-radius circular area centred on
12~00~00 +30~00~00 J2000, but could have done elsewhere with available SDSS,
USNO-B1, and 2MASS data.
The high galactic latitude of the field, $b \sim$ 78\,deg, minimises the
number of systematic errors of USNO-B1 proper motions in crowded fields.
Studying a larger survey area is feasible, but the data loading might slow down
the whole process. 
The common steps of the analysis have been as follow:

\begin{itemize}
\item With the Aladin {\tt Server selector}, load two images taken at different
epochs. 
Although one can manage without them, background images facilitate enormously
the source identification.
Instead of loading one of the {\tt Aladin images} (e.g. 1.7$\times$1.7\,deg$^2$
E-DSS1), I~recommend loading Digitized Sky Survey images from ESO (Garching)
in the {\tt DSS} button.
They take longer to be loaded, but their spatial resolution is highly improved.
Load, for example, 120$\times$120\,arcmin$^2$ DSS1 POSS1-Red/UKST-Blue and DSS2
POSS2UKSTU-Red images centred on the above coordinates. 
In this case, epochs of observations were J1955.275 (DSS1) and J1991.261 (DSS2).
\item Load all Simbad sources at 1\,deg to the central coordinates (button {\tt
Simbad} in {\tt Catalogs} column).
\item Load all 2MASS sources at 1\,deg to the central coordinates (button {\tt
Surveys} in {\tt Catalogs} column, pick up {\tt 2MASS}).
\item Load all USNO-B1 sources (button {\tt Surveys}, pick up {\tt USNO-B1}).
\item Load all SDSS sources (button {\tt Surveys}, pick up {\tt SDSS-DR6} --
note that it is the sixth SDSS data release).
They may take up to one minute to be loaded.
\item In the {\tt Tools} button, open the {\tt Catalog cross-match tool}.
Cross-match 2MASS and USNO-B1 (in this order) with the default threshold
in source separation (4\,arcsec) and match method (best matches).
\item Cross-match the previous output ({\tt XMatch result1}) with SDSS.
Afterwards, to save computer memory, one can delete planes in the Aladin plane
stack and leave only the DSS1 and DSS2 images, the Simbad sources (in red), and
the last output with the SDSS+USNO-B1+2MASS sources, {\tt XMatch result2} (in
cyan -- one can change the properties of the selected plane, such as colour or
shape, with the right mouse button).
\item Select all the sources in the {\tt XMatch result2} plane.
In the {\tt Tools} button, open {\tt VOplot} (the VO-India 2-D plotter for
selected objects).
For clarity, one can plot a proper motion diagram ({\tt pmDE} in Y axis, {\tt
pmRA} and active reversal, {\tt Rev}, in X axis).
\item In the {\tt VOplot} window, create a new data subset with the button {\tt
Create new filter}.
Enter data subset name (whatever: e.g. {\em Rosebud}) and condition.
I used the following condition: \begin{verbatim} 
sqrt($18 * $18 + $19 * $19) > 100} 
&& $5 < 15.5, 
\end{verbatim} which translates into $(\mu_\alpha^2 \cos^2{\delta} +
\mu_\delta^2)^{1/2} >$ 100\,mas\,a$^{-1}$ and $J <$ 15.5\,mag. 
The magnitude restriction is to minimise the number of faint sources with
spurious high proper motions (the great majority of high-proper motion star
candidates with $J >$ 15.5\,mag turn out to be USNO-B1 errors with $\mu
\approx 0$ when studied in detail -- e.g. by comparing DSS1 and DSS2 images). 
\item In the {\tt VOplot} window, select all the sources in the new data subset.
They are marked in the Aladin view window and are displayed in the measurement
window.
One can export the results in ascii to a file with the {\tt Aladin script
console} in the {\tt Tools} button or investigate each source in the measurement
window just with a few mouse clicks.
\end{itemize} 

There are 32 sources that satisfy the proper motion and $J$-band magnitude
requisites. 
After visual inspection, I~discarded two of them because of incorrect USNO-B1
proper motions: one is a Zwicky galaxy in a group of galaxies and the other one
is a background double source with null proper motion.
The remaining 30 sources have marked non-zero proper motions in the DSS1 and
DSS2 images (one can do a blink sequence with the button {\tt Image
associations} [{\tt asoc}] in the tool bar). 
I~also checked the consistency of proper motions of several targets as tabulated
by USNO-B1 and the SuperCOSMOS Science Archive (Hambly et~al. 2001).
Of the 30 objects, seven are bright stars and saturate or are in the non-linear
regime in at least the SDSS $i$ and $z$ images. 
The two brightest stars, HD~104379 (HIP~58615) and BD+30~2204, saturate at all
SDSS passbands. 
There is spectral type determination at F8--G5 for five of them (Schwassmann \&
van~Rhijn 1947).
The remaining two bright stars, G~121--36 and
2MASS~J120125.09+295057.0\footnote{It is discussed here for the first time.
Coordinates: 12~01~25.09 +29~50~57.0 J2000; 
proper motion: ($\mu_\alpha \cos{\delta}$, $\mu_\delta$) = (--110,
+12)\,mas\,a$^{-1}$; 
magnitudes: $g$ = 12.989$\pm$0.001\,mag, $J$ = 9.671$\pm$0.020\,mag.} seem to be
normal M~dwarfs.

\begin{table*}
  \begin{center}
    \caption{NVSS radio sources towards $\sigma$~Orionis with 2MASS
    near-infrared sources at less than 10\,arcsec.}\vspace{1em}
    \renewcommand{\arraystretch}{1.2}
    \begin{tabular}[h]{l ccc cc l}
      \hline
NVSS      	& $\alpha$      & $\delta$ 	& $S_\nu$(1.4\,GHz)& $\Delta \alpha$       & $\Delta \delta$	& Remarks			\\
 		& (J2000)	& (J2000)	& [mJy]		& [arcsec]	        & [arcsec]		&				\\
      \hline
054019--024403	& 05 40 19.87	& --02 44 03.5	&  4.9$\pm$0.9	& +1.9$\pm$1.2	        & --9$\pm$15		& In cluster of galaxies	\\		 
053727--024002	& 05 37 27.87	& --02 40 02.9	& 25.6$\pm$0.9	& --0.99$\pm$0.08	& +0.3$\pm$0.9		& 2E 1448			\\		 
053658--023854	& 05 36 58.71	& --02 38 54.6	&  2.8$\pm$0.5	& +4.0$\pm$0.4	        & --4$\pm$7		& Faint, blue, extended source	\\		 
053859--023409	& 05 38 59.00	& --02 34 09.8	&  3.2$\pm$0.5	& +2.4$\pm$0.4	        & +3$\pm$6		& Faint, extended source	\\ 
053911--023326	& 05 39 11.33	& --02 33 26.7	&  2.1$\pm$0.5	& --1.0$\pm$0.6	        & +6$\pm$10		& Mayrit 425070			\\		 
053755--023305	& 05 37 55.42	& --02 33 05.7	& 15.1$\pm$0.7	& --2.6$\pm$0.1	        & --0.4$\pm$1.3		& Mayrit 757283			\\		 
053957--022613	& 05 39 57.72	& --02 26 13.7	& 20.9$\pm$0.8	& +0.9$\pm$0.08	        & --5.4$\pm$1.0		& 2MASS J05395766--0226083	\\		 
053817--022455	& 05 38 17.45	& --02 24 55.9	&  5.3$\pm$0.6	& --3.0$\pm$0.3		& --7$\pm$6		& Faint, blue, extended source	\\		 
053904--021841	& 05 39 04.05	& --02 18 41.6	& 34.1$\pm$1.1	& +8.60$\pm$0.08	& +4.4$\pm$0.8		& In cluster of galaxies	\\
     \hline \\
      \end{tabular}
    \label{tab:radiosigmaorionis}
  \end{center}
\end{table*}
%

Next, I~focus on the 23 high proper-motion objects that do not saturate in the
SDSS images.
Of them, 17 are new and six had been previously detected by Giclas et~al.
(1965; one object), Luyten (1979; three), and L\'epine \& Shara (2005; two).
With the available proper motion and photometric data ($ugrizJHK_{\rm s}$),
I~have constructed several diagrams to ascertain the nature of the 23 objects.
In Fig.~\ref{fig:figure_su2}, I~show two of these diagrams: an SDSS-based
colour-colour diagram to the left, and the $g$-band reduced proper motion as a
function of the colour $g-J$ to the right.
Reduced proper motion is defined as $H_g \equiv g + 5 \log{\mu} +5$ ($\mu$ in
arcsec\,a$^{-1}$), while the colour $g-J$ is an indicator of effective
temperature. 
All the high proper motion objects except five have reduced proper motions and
SDSS-2MASS colours typical of M2--6 dwarfs in the field (L\'epine \& Shara
2005; West et~al. 2008).
The five exceptions are listed in Table~\ref{tab:SU2}.
The bluest outlier is GD~147, which was reported by Giclas et~al. (1965) as a
promising white dwarf suspect. 
Two previously unknown objects are located close to GD~147 in the two diagrams
in Fig.~\ref{fig:figure_su2}, and are classified in this proceeding as white
dwarf or blue subdwarf candidates.
For naming them, I~have followed the ``Albus'' ({\em white}) nomenclature
introduced by Caballero \& Solano (2007).
White dwarf GD~147 is more than two magnitudes brighter than Albus~2 and~3 in
the {\em GALEX} NUV (near-ultraviolet) passband, indicating a possible subdwarf
nature for the latters.
The other two outliers are LP~320--47 and a new object, for which I~have
followed the ``Ruber'' ({\em red}) nomenclature introduced by Caballero \&
Solano (2008). 
The latter star has red-optical and near-infrared colours typical of $\sim$M0
dwarfs, but redder colours at bluer wavelengths ($ugr$).
This fact and its red $g-J$ colour for its reduced proper motion may indicate a
subgiant phase.
Last, an abnormally low metallicity may be responsible of the peculiar optical
colours observed in LP~320--47.
A dedicated spectroscopic follow-up is needed to ascertain the nature of this
unusual object.

\section{The $\sigma$~Orionis cluster: A Space Odyssey}

The brightest star in the $\sigma$~Orionis cluster ($\tau \sim$ 3\,Ma, $d \sim$
385\,pc) is the eponymous $\sigma$~Ori multiple system, which is famous for
illuminating the Horsehead Nebula.
The cluster is the richest hunting ground for young low-mass brown dwarfs and
planetary-mass objects (substellar objects with theoretical masses below the
deuterium burning mass limit) and an important region for investigating
youth features in stars, such as X-ray emission, jets, photometric variability,
or discs (see a bibliographic review in Caballero 2008b).
However, the $\sigma$~Orionis stellar population was relatively poorly known a
few years ago. 

The Aladin-based cross-match between Tycho-2 and 2MASS in Caballero (2007b) was
the first effort to build a comprehensive list of bright stars in the
$\sigma$~Orionis cluster. 
Membership status of each star was based on optical and near-infrared colours
and magnitudes, proper motions, {\em IRAS} infrared excesses, X-ray emission,
and spectroscopic data from the literature. 
Collateral results in this work were the first determination of the mass
function and the disc frequency in the high mass domain of the cluster and the
astrometric confirmation of overlapping of different young star populations
in the Orion~Belt.

In Caballero (2008b), I~extended the catalogue of cluster members, candidate
members, and non-members towards the limit imposed by the $i$-band depth of the
Deep Near Infrared Survey of the Southern Sky (DENIS). 
The basis of the work was an optical/near-infrared correlation between the
2MASS and DENIS catalogues in a circular area of radius 30\,arcmin centred on
$\sigma$~Ori~AB.
The analysis was supported by an exhaustive bibliographic search of confirmed
cluster members with signposts of youth and by additional X-ray, mid-infrared,
and astrometric data.
The output of the search, the Mayrit catalogue of $\sigma$~Orionis stars and
brown dwarfs, has turned out to be a very useful tool for studying the spatial
distribution (Caballero 2008a; Bouy et~al. 2008), disc frequency as a function
of mass (Luhman et~al. 2008), initial mass function and multiplicity (Caballero
2008c; N. Lodieu et~al., in~prep.), X-ray emission (L\'opez-Santiago \&
Caballero 2008; Caballero et~al. 2009; E.~Francisoni et~al., in~prep.), and
spectroscopic properties of cluster low-mass stars and brown dwarfs (Caballero
et~al. 2008a; A.-M. Cody et~al., in~prep.). 
The preparation of an updated version of the Mayrit catalogue, with UKIDSS
photometry, spectral types, radial velocities, and Li~{\sc i} $\lambda$670.8\,nm
pseudo-equivalent widths, is on-going (Caballero 2008c).

\subsection{Radio sources towards $\sigma$~Orionis}

\begin{figure*}
\centering
\includegraphics[width=0.45\linewidth]{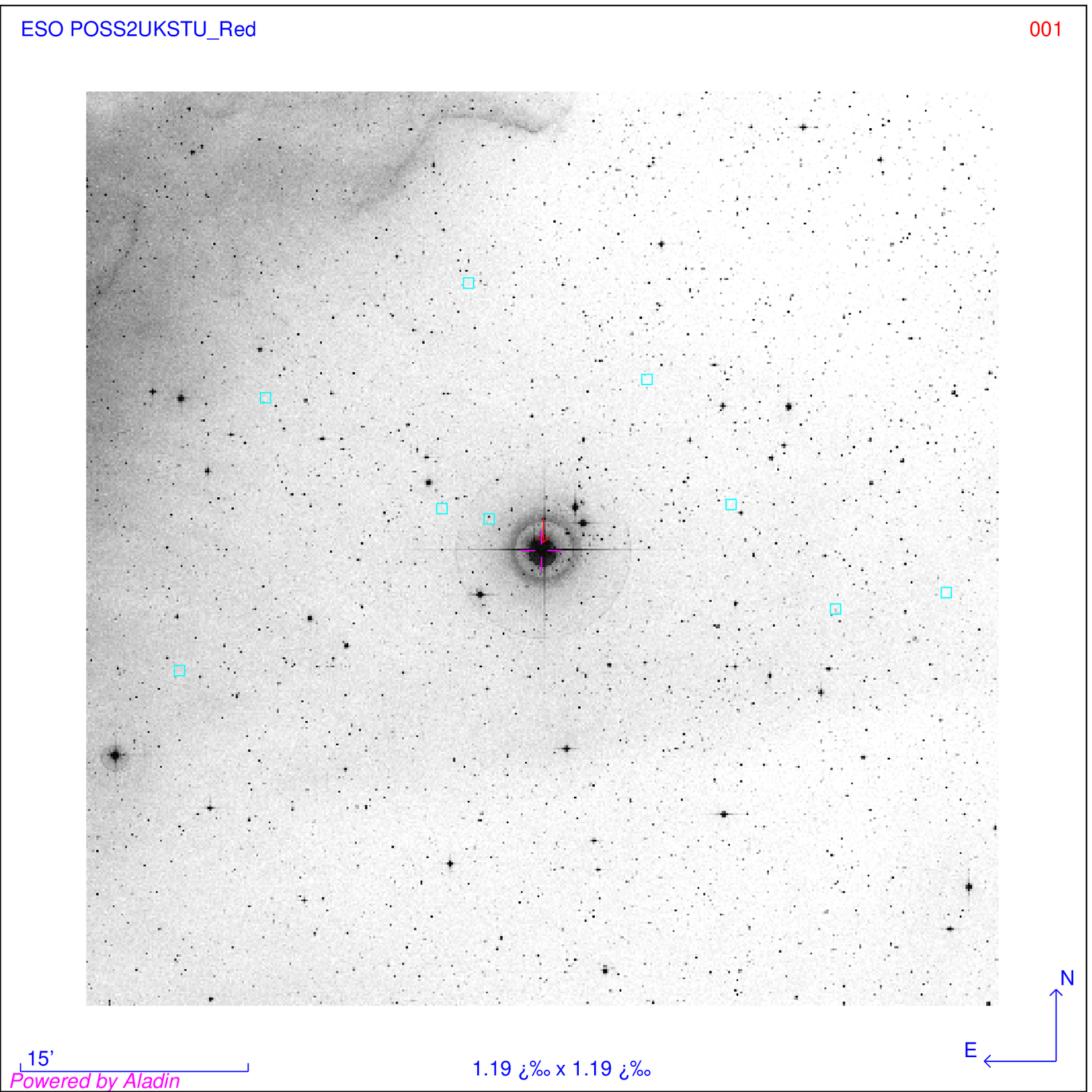}
\includegraphics[width=0.45\linewidth]{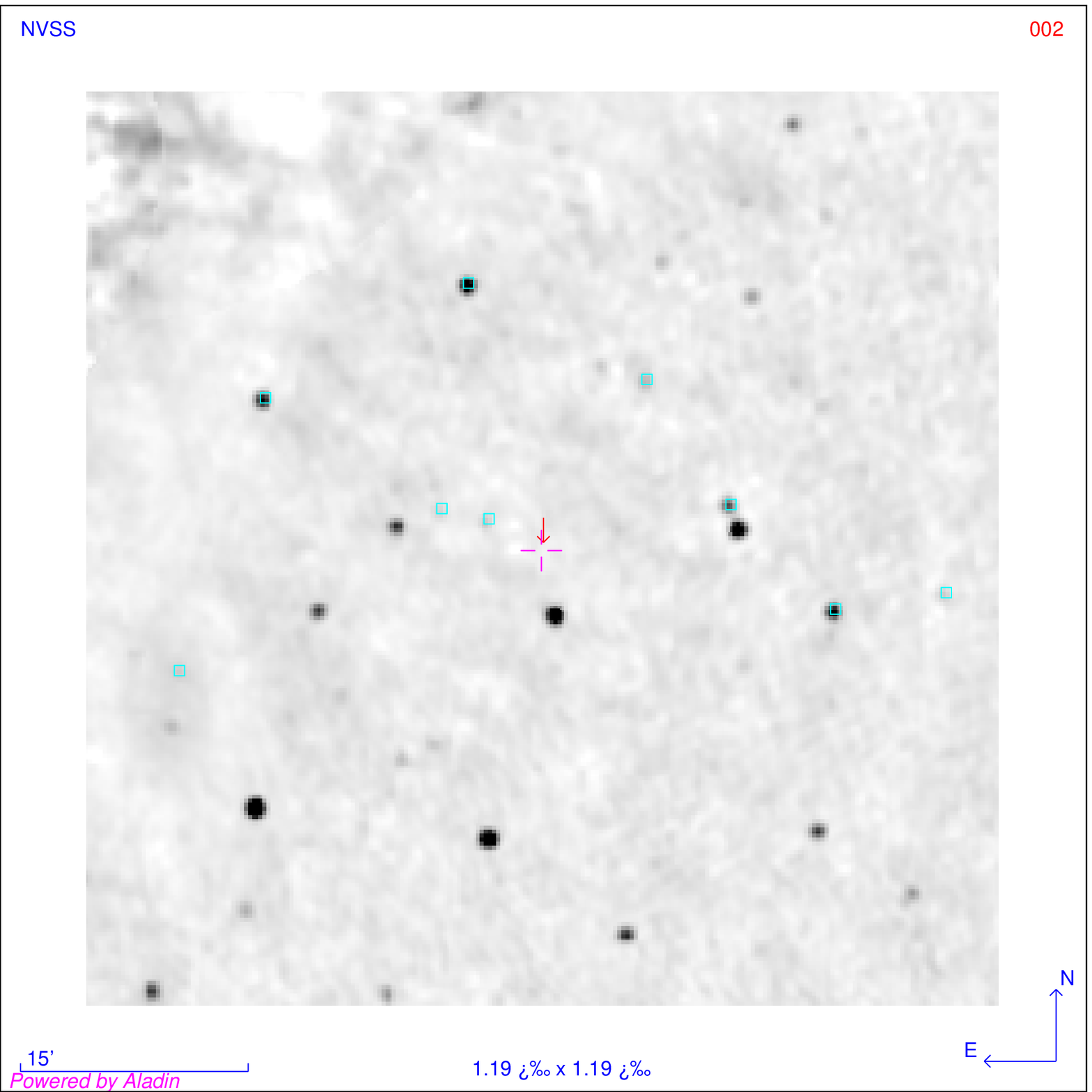}
\caption{ESO POSSII UKST Red ({\em left}) and NVSS ({\em right}) images centred
on $\sigma$~Ori~AB (marked with a cross). 
NVSS radio sources with 2MASS near-infrared sources at less than 10\,arcsec are
marked with (cyan) squares.
Size of the field of view is 1.0\,deg. 
North is up and east is left.
\label{fig:figure_eps_opticalradio}}
\end{figure*}

In the course of radio-continuum observations of a small sample of hot stars,
Drake (1990) detected a faint (probably non-thermal) radio emission at 2, 6, and
20\,cm coming from the pair $\sigma$~Ori~AB in the centre of the
$\sigma$~Orionis cluster. 
He found a significant radio to optical positional offset of 2.1\,arcsec.
The explanation for this discrepancy came when van~Loon \& Oliveira (2003) found
a mid-infrared source that is a proto-planetary disc being dispersed by the
intense ultraviolet radiation from $\sigma$~Ori~AB.
The X-ray and near-infrared counterparts of the star in the densest part of the
dust cloud was afterwards discovered by Sanz-Forcada et~al. (2004) and Caballero
(2005), respectively.
This star, known as $\sigma$~Ori~IRS1 (Mayrit~3020~AB), is actually double (Bouy
et~al. 2008). 
The free-free radio emission arises from a photo-ionized region at the interface
between the radiation field of $\sigma$~Ori~AB and the dust cloud.
Previously, Drake et~al. (1987) had found non-thermal radio emission from
$\sigma$~Ori~E, a magnetic, B2Vp, He-rich star at only 42\,arcsec to
$\sigma$~Ori~AB.
In this case, the emitting volume is non-spherical and co-rotates with the star
($P$ = 1.19\,d; Leone \& Umana 1993).
Density fluxes at centimetric wavelengths of both $\sigma$~Ori~AB and E are
relatively low, of about a few millijanskys (Wendker 1995).
Finding stronger radio emitters in the cluster would ease the determination of
its parallactic distance via very long baseline interferometry {\em \`a la}
Orion Nebula Cluster (Sandstrom et~al. 2007; Menten et~al. 2007).
Unfortunately, no third cluster radio emitter is known to date.
Drake (1990) reported another three radio sources in the vicinity of
$\sigma$~Ori~AB ([D90]~1, 2, and~3), while Caballero et~al. (2007) presented a
faint, red, near-infrared source at 1.5$\pm$1.0\,arcsec to another relatively
bright radio source (TXS~0537--029).
The four sources seem to have extragalactic nature, especially the latter, for
which Caballero et~al. (2007) identified the possible host galaxy.

In the second VO practical example, I~have used Aladin to look for the possible
2MASS counterparts of the sources in the 1.4\,GHz National Radio Astronomy
Observatory Very Large Array Sky Survey (NVSS; Condon et~al. 1998) at less than
30\,arcmin to $\sigma$~Ori~AB.
I~carried out a methodology similar to that presented in
Section~\ref{section:su2}, but cross-matching the NVSS and 2MASS catalogues
using a threshold in source separation of 10\,arcsec.
Fig.~\ref{fig:figure_eps_opticalradio} illustrates this search.
The nine cross-matched NVSS radio sources are listed in
Table~\ref{tab:radiosigmaorionis}. 
Among the non-cross-matched NVSS sources, there are two radio
sources discussed above ([D90]~1 and TXS~0537--029).
Of the cross-matched sources, three are associated to faint extended sources
with a likely extragalactic nature.
Another two radio sources are surrounded by a number of faint blue optical
sources with USNO-B1/2MASS colours typical of unresolved galaxies, which might
form clusters of galaxies. 
In these cases, it is difficult to ascertain which is the actual origin of the
radio emission.
The second brightest NVSS source in Table~\ref{tab:radiosigmaorionis} is
associated to a known galaxy with hard X-ray emission and a power-law photon
index consistent with it being a Seyfert galaxy with a supermassive black hole
(2E~1448; L\'opez-Santiago \& Caballero 2008).
There remain three NVSS radio sources.
In two cases, the closest 2MASS source is a $\sigma$~Orionis cluster member in
the Mayrit catalogue.
In the third case, a Mayrit star is the second closest 2MASS source, only
a few arcseconds further than the one listed in
Table~\ref{tab:radiosigmaorionis}. 
Below, I~enumerate the three of them:

\begin{itemize}
\item NVSS~053911--023326, at $\rho$ = 6$\pm$10\,arcsec to Mayrit~425070 (S\,Ori
J053911.4--023333).
The 2MASS possible counterpart is a very low-mass cluster star with Li~{\sc i}
in absorption and low gravity spectral features, but it lacks a disc (Caballero
2008b and references therein). 
No H~{\sc i} 21\,cm line is expected to come from this object; it may come from
an unidentified background galaxy instead.
\item NVSS~053957-022613, at $\rho$ = 5.5$\pm$1.0\,arcsec to 2MASS
J05395766--0226083 and $\rho \sim$ 9.4\,arcsec to the confirmed cluster member
Mayrit~1245062 (S\,Ori J053958.1--022619). 
Despite its Li~{\sc i} absorption and moderate H$\alpha$ emission, the accretion
disc surrounding Mayrit~1245062 cannot be dense and hot enough to emit with a
density flux of more than 20\,mJy.
\item NVSS~053755-023305, at $\rho$ = 2.7$\pm$1.3\,arcsec to Mayrit~757283
(S\,Ori~35).
On the contrary to the other two stars, the disc surrounding Mayrit~757283 is
apparent in {\em Spitzer} observations at 8\,$\mu$m (Hern\'andez et~al. 2007).
However, the possible two-lobe 1.4\,GHz emission, the significant positional
offset, and the intensity of the radio emission may indicate a possible
extragalactic nature.
\end{itemize}

To sum up, $\sigma$~Ori~AB+IRS1 and $\sigma$~Ori~E remain as the only known
radio emitters {\em in} the $\sigma$~Orionis cluster.

\section{New hunting grounds for brown dwarfs}

Numerous populations of brown dwarfs and planetary mass-objects have been found
in other young star-forming regions different from the $\sigma$~Orionis cluster,
such as $\rho$~Ophiuchi, Chamaeleon~I, or the Orion Nebula Cluster, but not in
so large quantities.
Besides, the much lower extinction in $\sigma$~Orionis facilitates the
spectro-photometric follow-up for confirmation of cluster membership.
Very few $\sigma$~Orionis analogues ($d <$ 0.5\,kpc, $\tau <$ 10\,Ma, $A_V <$
1\,mag) do exist, like the Collinder~69 cluster around $\lambda$~Ori in the
Orion Head.

Two of these $\sigma$~Orionis analogues were identified and characterised for
the first time by Caballero \& Solano (2008).
They catalogued about 500 young stars and candidates surrounding the OB
supergiants Alnilam ($\epsilon$~Ori -- in the Collinder~70 cluster) and Mintaka
($\delta$~Ori) in the Orion Belt. 
Procedures and membership selection criteria, based on cross-matches of the
Tycho-2, DENIS, and 2MASS catalogues with Aladin and construction of
colour-magnitude and proper-motion diagrams, were very similar to those used by
Caballero (2007b, 2008b) in the Mayrit catalogue.
Many new detections of stars with near- and mid-infrared excesses due to
circumstellar discs and X-ray emission arised from the survey\footnote{In the
course of this survey, Caballero \& Solano (2007) serendipitously identified
Albus~1 (CPD--20~1123), one of the brightest He-B subdwarfs yet found (Vennes
et~al. 2007).}.

Except Taurus, all star-forming regions where young brown dwarfs have been found
have in common the presence of early-type stars that ionise the interstellar
medium (e.g. {\em all} clusters above plus NGC~2264, IC~2391, Pleiades...).
Therefore, looking for new sites for substellar searches is synonymous with
looking  for agglomerates of early-type stars.
Following this idea, Caballero \& Dinis (2008) studied the spatial structure and
sub-structure of regions rich in {\em Hipparcos} stars with blue $B_T-V_T$
colours.
The exhaustive all-sky analysis of the membership in agglomerate of the 406
selected stars was carried out with Aladin. 
Most of the 35 identified agglomerates were associated to previously known
clusters and OB associations.
Brown dwarfs have been searched for in a significant fraction of these
agglomerates (e.g. Orion Nebula Cluster, Pleiades).
We listed seven agglomerates (including NGC~2451A, vdBH~23, Trumpler~10, and the
new, nearby, young, open cluster P~Puppis) as new sites for substellar searches.
The brown dwarf populations of some of these agglomerates await discovery and
are suitable to be analysed with Aladin.

\subsection{X-ray sources in Collinder~70}

\begin{table*}
  \begin{center}
    \caption{X-ray T~Tauri stars surrounding Alnilam with colours $J-K_{\rm s}
    >$ 1.2\,mag.}\vspace{1em}
    \renewcommand{\arraystretch}{1.2}
    \begin{tabular}[h]{lcccccc}
      \hline
Name      		& $\alpha$      	& $\delta$ 	& $\rho$ 	& $J$			& $K_{\rm s}$		& $F_X$			\\
 	    		& (J2000)		& (J2000)	& [arcsec] 	& [mag]			& [mag]			& [10$^{-17}$\,W\,m$^{-2}$]\\
      \hline
Kiso A--0904 41       	& 05 35 53.49	       	& --01 23 04.4  & 0.94	       & 12.65$\pm$0.02        & 11.43$\pm$0.02        & 4$\pm$3	       \\
StHA 47	       		& 05 35 22.93	       	& --01 11 24.3  & 0.19	       & 10.457$\pm$0.019      & 9.18$\pm$0.02         & 18$\pm$4	       \\
V583 Ori	       	& 05 36 35.15	       	& --01 02 16.8  & 0.86	       & 12.12$\pm$0.02        & 10.881$\pm$0.019      & 4$\pm$3	       \\
      \hline \\
      \end{tabular}
    \label{tab:xrayalnilam}
  \end{center}
\end{table*}

Several young stars and candidates surrounding Alnilam and Mintaka in the
survey by Caballero \& Solano (2008) are X-ray emitters detected by the {\em
Einstein}, {\em ROSAT}, and {\em XMM-Newton} space missions.
Nevertheless, the great majority of them had never been reported in the
literature.  
In the third and last VO practical example, I~complement that survey with the
identification of the optical and near-infrared counterparts of X-ray sources in
the {\em XMM-Newton} Serendipitous Source Catalogue (2XMM) surrounding Alnilam.

\begin{figure}
\centering
\includegraphics[width=1.0\linewidth]{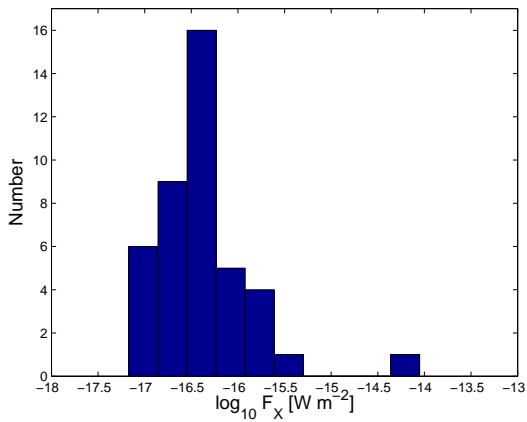}
\caption{X-ray ``luminosity function'' of young stars in the Alnilam region.
Fluxes and luminosities are related through a $4 \pi d^2$ factor (where $d \sim$
0.4\,kpc is the heliocentric distance of Collinder~70).
\label{fig:figure_eps_alnilamX_1}}
\end{figure}

\begin{figure}
\centering
\includegraphics[width=1.0\linewidth]{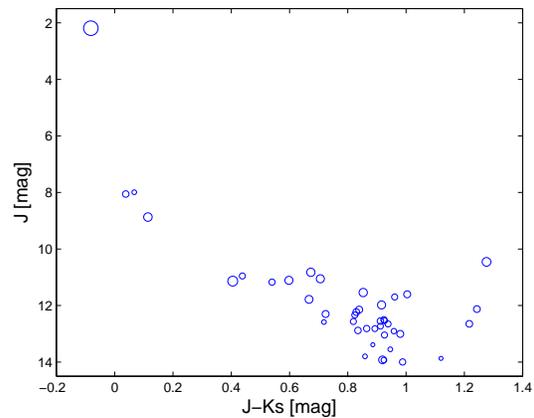}
\caption{Near-infrared colour-magnitude diagram of the X-ray young stars in the
Alnilam region.
Note the three stars with colours $J-K_{\rm s} >$ 1.2\,mag.
\label{fig:figure_eps_alnilamX_2}}
\end{figure}

Again, the basis of this analysis was an Aladin-based cross-match between 2XMM
and 2MASS catalogues in the core of the Collinder~70 cluster.
Of the 78 2XMM sources in the area, 52 have a 2MASS source within a threshold in
source separation of 10\,arcsec.
Among them, one is an active F4V star in the foreground and nine are X-ray
galaxies. 
The extragalactic nature of some of the latter objects is patent from 2MASS
photometry, DSS images, or alternative multi-wavelength data (e.g. the
radio-galaxy NVSS~J053627--005937).
There remain 42 sources with magnitudes and colours consistent with membership
in young stellar populations in the Orion Belt, of which only eleven were
previously known:
four are bright Henry Draper stars (including Alnilam itself), three are known
T~Tauri stars (see below), two are {\em Einstein} stars catalogued by Caballero
\& Solano (2008), one is a photometric cluster member candidate with periodic
photometric variability ([SE2005]~118, $P$ = 98$\pm$5\,h, A$_I$ = 0.089\,mag;
Scholz \& Eisl\"offel 2005), and the last one is associated to X-ray source
1WGA~J0537.1--0109, which is tabulated as variable by 2XMM.
Most of the 42 X-ray detections are new. 
Except in one case, the separation between the 2XMM and 2MASS coordinates is
less than or about 3\,arcsec.
Figs.~\ref{fig:figure_eps_alnilamX_1} and~\ref{fig:figure_eps_alnilamX_2}
illustrate the X-ray search.

The three X-ray T~Tauri stars, listed in Table~\ref{tab:xrayalnilam}, are the
reddest young stars in my sample and have colours $J-K_{\rm s} >$ 1.2\,mag.
This fact, together with the detection of the H$\alpha$ line in strong emission
(Haro \& Moreno 1953; Stephenson 1986; Wiramihardja et~al. 1989 -- V583~Ori is
also a known photometric variable), supports the disc scenario.
The X-ray emission from Kiso~A--0904~41 and V583~Ori is faint, as expected in
T~Tauri stars with absorbing discs (e.g. Neuh\"auser et~al. 1995).
However, the disc around StHA~47, which has a stronger X-ray emission, may be
face-on (see some examples of face-on discs in $\sigma$~Orionis in
L\'opez-Santiago \& Caballero 2008).
All these detections deserve a careful analysis of the original {\em XMM-Newton}
data that will be presented in a forthcoming publication.

\section{The future is bright!}

With Aladin and a little of imagination, one can do a lot of interesting
science. 
However, I~have only a vague idea of what one will be able to do when the data
from future deep, multi-wavelength, all-sky surveys will be available through
Aladin. 
The next generation of ground-based surveys is here:
RAVE\footnote{\tt http://www.rave-survey.aip.de},
UKIDSS\footnote{\tt http://www.ukidss.org},
SDSS-III\footnote{\tt http://www.sdss3.org},
Pan-STARRS\footnote{\tt http://pan-starrs.ifa.hawaii.edu}, and
LSST\footnote{\tt http://www.lsst.org}.
But the ``virtual'' revolution will come with next all-sky space missions, such
as NASA's {\em WISE}\footnote{\tt http://wise.ssl.berkeley.edu} (launch
November 2009) and, especially, ESA's {\em GAIA}\footnote{\tt
http://gaia.esa.int} (launch December 2011).
Do not keep still in the mean time!


\section*{Acknowledgments}

JAC is an {\em investigador Juan de la Cierva} at the UCM.
Financial support was provided by the Universidad Complutense de Madrid,
the Comunidad Aut\'onoma de Madrid, the Spanish Ministerio Educaci\'on y
Ciencia, and the European Social Fund.  
This research has made use of the SIMBAD, operated at Centre de Donn\'ees
astronomiques de Strasbourg, France, and the NASA's Astrophysics Data System.


\end{document}